\documentclass[onecolumn,showpacs,preprintnumbers,amsmath,amssymb,eqsecnum]{revtex4}
\usepackage{dcolumn}
\usepackage{bm}
\usepackage[dvips]{graphicx}
\usepackage{wrapfig}
\usepackage{psfrag}
\usepackage{color}
\begin{document}


\def\sh{\mathop{\rm sh}\nolimits}
\def\ch{\mathop{\rm ch}\nolimits}
\def\var{\mathop{\rm var}}\def\exp{\mathop{\rm exp}\nolimits}
\def\Re{\mathop{\rm Re}\nolimits}
\def\Sp{\mathop{\rm Sp}\nolimits}
\def\kp{\mathop{\text{\ae}}\nolimits}
\def\bk{{\bf {k}}}
\def\bp{{\bf {p}}}
\def\bq{{\bf {q}}}
\def\lra{\mathop{\longrightarrow}}
\def\Const{\mathop{\rm Const}\nolimits}
\def\sh{\mathop{\rm sh}\nolimits}
\def\ch{\mathop{\rm ch}\nolimits}
\def\var{\mathop{\rm var}}
\def\mK{\mathop{{\mathfrak {K}}}\nolimits}
\def\mR{\mathop{{\mathfrak {R}}}\nolimits}
\def\mv{\mathop{{\mathfrak {v}}}\nolimits}
\def\mV{\mathop{{\mathfrak {V}}}\nolimits}
\def\mD{\mathop{{\mathfrak {D}}}\nolimits}
\def\mN{\mathop{{\mathfrak {N}}}\nolimits}
\def\mS{\mathop{{\mathfrak {S}}}\nolimits}
\def\mT{\mathop{{\mathfrak {T}}}\nolimits}
\def\mt{\mathop{{\mathfrak {t}}}\nolimits}
\def\me{\mathop{{\mathfrak {e}}}\nolimits}
\def\mE{\mathop{{\mathfrak {E}}}\nolimits}

\newcommand\ve[1]{{\mathbf{#1}}}

\def\Re{\mbox {Re}}
\newcommand{\Z}{\mathbb{Z}}
\newcommand{\R}{\mathbb{R}}
\def\mK{\mathop{{\mathfrak {K}}}\nolimits}
\def\mk{\mathop{{\mathfrak {k}}}\nolimits}
\def\mR{\mathop{{\mathfrak {R}}}\nolimits}
\def\mv{\mathop{{\mathfrak {v}}}\nolimits}
\def\mV{\mathop{{\mathfrak {V}}}\nolimits}
\def\mD{\mathop{{\mathfrak {D}}}\nolimits}
\def\mN{\mathop{{\mathfrak {N}}}\nolimits}
\def\ml{\mathop{{\mathfrak {l}}}\nolimits}
\def\mf{\mathop{{\mathfrak {f}}}\nolimits}
\newcommand{\ccm}{{\cal M}}
\newcommand{\cE}{{\cal E}}
\newcommand{\cV}{{\cal V}}
\newcommand{\cI}{{\cal I}}
\newcommand{\cR}{{\cal R}}
\newcommand{\cK}{{\cal K}}
\newcommand{\cH}{{\cal H}}
\newcommand{\cW}{{\cal W}}

\def\br{\mathop{{\bf {r}}}\nolimits}
\def\bS{\mathop{{\bf {S}}}\nolimits}
\def\bA{\mathop{{\bf {A}}}\nolimits}
\def\bJ{\mathop{{\bf {J}}}\nolimits}
\def\bn{\mathop{{\bf {n}}}\nolimits}
\def\bg{\mathop{{\bf {g}}}\nolimits}
\def\bv{\mathop{{\bf {v}}}\nolimits}
\def\be{\mathop{{\bf {e}}}\nolimits}
\def\bp{\mathop{{\bf {p}}}\nolimits}
\def\bz{\mathop{{\bf {z}}}\nolimits}
\def\bbf{\mathop{{\bf {f}}}\nolimits}
\def\bb{\mathop{{\bf {b}}}\nolimits}
\def\ba{\mathop{{\bf {a}}}\nolimits}
\def\bx{\mathop{{\bf {x}}}\nolimits}
\def\by{\mathop{{\bf {y}}}\nolimits}
\def\br{\mathop{{\bf {r}}}\nolimits}
\def\bs{\mathop{{\bf {s}}}\nolimits}
\def\bH{\mathop{{\bf {H}}}\nolimits}
\def\bk{\mathop{{\bf {k}}}\nolimits}
\def\be{\mathop{{\bf {e}}}\nolimits}
\def\bnul{\mathop{{\bf {0}}}\nolimits}
\def\bq{{\bf {q}}}
\def\diag{\mathop{\rm diag}\nolimits}

\newcommand{\oV}{\overline{V}}
\newcommand{\vkp}{\varkappa}
\newcommand{\os}{\overline{s}}
\newcommand{\opsi}{\overline{\psi}}
\newcommand{\ov}{\overline{v}}
\newcommand{\oW}{\overline{W}}
\newcommand{\oPhi}{\overline{\Phi}}

\def\mI{\mathop{{\mathfrak {I}}}\nolimits}
\def\mA{\mathop{{\mathfrak {A}}}\nolimits}

\def\st{\mathop{\rm st}\nolimits}
\def\tr{\mathop{\rm tr}\nolimits}
\def\sign{\mathop{\rm sign}\nolimits}
\def\d{\mathop{\rm d}\nolimits}
\def\const{\mathop{\rm const}\nolimits}
\def\O{\mathop{\rm O}\nolimits}
\def\Spin{\mathop{\rm Spin}\nolimits}
\def\exp{\mathop{\rm exp}\nolimits}
\def\SU{\mathop{\rm SU}\nolimits}
\def\mU{\mathop{{\mathfrak {U}}}\nolimits}
\newcommand{\cU}{{\cal U}}
\newcommand{\cD}{{\cal D}}

\def\curl{\mathop{\mathrm{curl}}\nolimits}
\def\mI{\mathop{{\mathfrak {I}}}\nolimits}
\def\mA{\mathop{{\mathfrak {A}}}\nolimits}
\def\mU{\mathop{{\mathfrak {U}}}\nolimits}

\def\st{\mathop{\rm st}\nolimits}
\def\tr{\mathop{\rm tr}\nolimits}
\def\sign{\mathop{\rm sign}\nolimits}
\def\d{\mathop{\rm d}\nolimits}
\def\const{\mathop{\rm const}\nolimits}
\def\O{\mathop{\rm O}\nolimits}
\def\Spin{\mathop{\rm Spin}\nolimits}
\def\exp{\mathop{\rm exp}\nolimits}

\title{Self-consistent equation for torsion arising as a consequence  of the Dirac sea quantum fluctuations in external classical electromagnetic and gravitational fields}

\author {S.N. Vergeles\vspace*{4mm}\footnote{{e-mail:vergeles@itp.ac.ru}}}

\affiliation{Landau Institute for Theoretical Physics,
Russian Academy of Sciences,
Chernogolovka, Moscow region, 142432 Russia \linebreak
and   \linebreak
Moscow Institute of Physics and Technology, Department
of Theoretical Physics, Dolgoprudnyj, Moskow region,
141707 Russia}

\begin{abstract} The quantum fluctuations of the Dirac field in external classical gravitational and electromagnetic fields are studied. A self-consistent equation for torsion is calculated, which is obtained using one-loop fermion diagrams.
\end{abstract}


\maketitle

\section{Introduction}

In this paper, we study the theory of gravity, minimally related to the massive Dirac field, formulated in the form of Cartan-Palatini. The Dirac field is considered to be quantized, while the gravitational (and electromagnetic) fields are classical.
Since the gauge fields are classical, the mean values of bilinear forms constructed using Dirac fields are exhausted by one-loop diagrams. The torsion tensor appears to be one of such  mean value.
This tensor appears as the fermion part of one of the equations of motion
(see the next Section, Eq. (\ref{Torsion})):
\begin{gather}
e^{\mu}_be^{\nu}_c\left({\cal D}_{\mu}e^a_{\nu}-{\cal D}_{\nu}e^a_{\mu}\right)
=i\frac{l_P^2}{4}\overline{\Psi}(\gamma^a\sigma_{bc}+\sigma_{bc}\gamma^a)\Psi
\label{Eq}
\end{gather}
By definition, the left-hand side of the last equation is the torsion tensor $\mT_{bc}^a$.

In this work, the mean value of the right-hand side of Eq. (\ref{Eq})  is calculated explicitly. Only divergent contributions (in momentum space) are taken into account. Therefore, the result can depend only on local geometric quantities (curvature and torsion tensors and their covariant derivatives). As a result of these calculations, the right-hand side of  Eq. (\ref{Eq}) turns out to be a local function of the curvature and torsion tensors and their covariant derivatives.
If we assume that the curvature tensor is given, then in this way a self-consistent equation for the torsion tensor arises. This equation is one of the equations of motion in the studied model.

The  self-consistent equation for torsion tensor (\ref{Self-consistent_Eq}) may be interesting for the following reasons. Even a superficial study of this equation shows that in the present cosmological epoch the torsion tensor generated by the Dirac field must be zero. This fact is consistent with experimental data.
However, conditions near the Big Bang may occur when the torsion tensor turns out to be nonzero.
To study the cosmological consequences arising from this, separate studies are needed.

In this paper, the question of the contribution to the right-hand side of Eq. (\ref{Eq}) from the Weyl fermion fields (neutrinos) remained open. This contribution must be calculated since Weyl fermions exist.
However, the corresponding computational procedure must be somewhat modified, since the Weyl fields are either massless or their masses are extremely small.

There is a good reason to consider gravitational fields as classical.
The quantum theory of gravity is a non-renormalizable theory. In particular, this means that quantum fluctuations of gravitational dynamical variables (tetrads and connections) are large on ultra-small scales of the order of the Planck length \cite{vergeles2021note}. But on the scales much larger than the Planck scale, these fluctuations decrease rapidly (according to a power law). Therefore, we will assume that when considering physics on scales that are much larger than Planck's, fluctuations of the gravitational degrees of freedom are insignificant, that is, these degrees of freedom are described by classical fields. That's why the gravitational fields $e^a_{\mu}$ and $\omega^{ab}_{\mu}$ are considered as classical. However, the quantum fluctuations of the Dirac field are important for wavelengths shorter than the Compton wavelength of the Dirac particle.

Recently, the problems of the dynamics of the Dirac and Weyl spin current in gravitational and electromagnetic fields have been actively studied. For example, in the
work \cite{de2021master} the master equations governing
the interaction between gravitational fields and gauge
invariant spin-current is derived. Note that the canonical spin current (\ref{Spin_Current}) is proportional to the torsion tensor, and therefore such a spin current in an external electromagnetic field is considered here.
In the paper \cite{chu2021induced} a variant of the spin current constructed with the help of the spin operator in the  Pauli-Lubanski form  is studied. In this paper, it is shown that the vacuum expectation value of such a spin in an external electromagnetic field is nonzero. It is shown below that the same vacuum expectation value of the spin current (\ref{Spin_Current}) is zero.
We refer the reader for the interesting results and a significant number of citations  on this topic to the review \cite{cranganore2021sugawara}.
The one-loop calculation of the effective action arising due to integration over the Dirac field in the same model (\ref{Dirac_Actin}) is contained in \cite{nascimento2022induced}. 

We also point to the review \cite{hehl1976general}, entitled "General relativity with spin and torsion",
which may be useful in getting acquainted with the problem under study here.

This article is structured as follows. Section II defines the model under study and introduces notation. Section III presents the results of calculations and writes out a self-consistent equation for the torsion tensor. Then comes the Conclusion.

\section{Definition of the model and technical means}

Let us write out the action of gravity coupled with a massive Dirac field in external classical electromagnetic and gravitational fields:
\begin{gather}
{\mA}={\mA}_g+{\mA}_{\Psi},
\nonumber \\
{\mA}_g=-\frac{1}{4l_P^2}\varepsilon_{abcd}\varepsilon^{\mu\nu\lambda\rho}\int
\mR^{ab}_{\mu\nu}e^c_{\lambda}e^d_{\rho}\d^4x,
\nonumber \\
\mR^{ab}_{\mu\nu}=\partial_{\mu}\omega^{ab}_{\nu}-\partial_{\nu}\omega^{ab}_{\mu}+
\omega^a_{c\mu}\omega^{cb}_{\nu}-\omega^a_{c\nu}\omega^{cb}_{\mu}=-\mR^{ba}_{\mu\nu},
\nonumber \\
{\mA}_{\Psi}=\int\d^4x\sqrt{-g}
\left\{\frac{i}{2}\tilde{e}^{\mu}_a
\left(\overline{\Psi}\gamma^a{\cal D}_{\mu}\Psi-\overline{{\cal D}_{\mu}\Psi}\gamma^a\Psi\right)
-m\overline{\Psi}\Psi\right\}.
\label{Dirac_Actin}
\end{gather}
Here
\begin{gather}
{\cal D}_{\mu}\Psi=\left(\partial_{\mu}-ieA_{\mu}+\omega_{\mu}\right)\Psi,
\nonumber \\
\omega_{\mu}
\equiv\frac12\omega_{ab\mu}\sigma^{ab},  \quad \sigma^{ab}=\frac14[\gamma^a,\gamma^b],
\label{Notations}
\end{gather}
The vector fields $\{\tilde{e}^{\mu}_a\},\,a=0,1,2,3$ form an local orthonormal basis, so that
\begin{gather}
g_{\mu\nu}\tilde{e}^{\mu}_a\tilde{e}^{\nu}_b=\eta_{ab}=\diag(1,-1,-1,-1),
\label{ONB}
\end{gather}
and $\tilde{e}^{\mu}_ae^b_{\mu}=\delta^b_a$.
Levi-Civita symbols are equal to units if their indices are ordered as $(0123)$.

We regard the lattice theory of gravity \cite{vergeles2021note} as a regularization of the continuum theory (\ref{Dirac_Actin}). Thus, divergent integrals in the calculation of fermionic loops should be cut off at lattice (or Planck) scales.

Equation $\delta{\mA}/\delta\omega^{ab}_{\mu}=0$ gives the definition of torsion:
\begin{gather}
\mT^a_{\mu\nu}\equiv {\cal D}_{\mu}e^a_{\nu}-{\cal D}_{\nu}e^a_{\mu}=
\frac{il^2_P}{4}\overline{\Psi}\left(\gamma^a\sigma_{bc}+\sigma_{bc}\gamma^a\right)\Psi e^b_{\mu}e^c_{\nu}.
\label{Torsion}
\end{gather}
Or, equivalently
\begin{gather}
\mT^a_{bc}=\frac{l^2_P}{4}\varepsilon^a_{\,\,\,bcd}\overline{\Psi}\gamma^5\gamma^d\Psi.
\label{Torsion_Defin}
\end{gather}
In the process of transforming the right side of Eq. (\ref{Torsion}) we used the equality
\begin{gather}
\gamma_a\sigma_{bc}+\sigma_{bc}\gamma_a=-i\varepsilon_{abcd}\gamma^5\gamma^d.
\nonumber
\end{gather}
Variation of action ${\mA}_{\Psi}$ with respect to connection $\omega_{\mu}^{bc}$ determines the expression for the Dirac spin current ${\mS}^a_{bc}$:
\begin{gather}
\delta_{\omega}{\mA}_{\Psi}\equiv-\frac12\int\d^4x\sqrt{-g}{\mS}^{\mu}_{bc}\delta\omega_{\mu}^{bc},
\nonumber \\
{\mS}^a_{bc}=-\frac12\varepsilon^a_{\,\,\,bcd}\overline{\Psi}\gamma^5\gamma^d\Psi.
\label{Spin_Current}
\end{gather}
Comparing (\ref{Torsion_Defin}) and (\ref{Spin_Current}), we find:
\begin{gather}
\mT^a_{bc}=-\frac{l^2_P}{2}\mS^a_{bc}.
\label{Torsion_sim_Spin.Current}
\end{gather}
Thus, the torsion tensor and the Dirac spin current are proportional in the considered model.

Hereinafter, $\langle...\rangle$ means averaging over the fermionic vacuum at fixed external classical  electromagnetic and gravitational fields.

The Dirac causal propagator is usually denoted as $S_c(x,y)$.
However, in order to avoid misunderstandings and confusion with the lower Latin indices of numerous tensors that are present in the formulas along with the causal propagator, we will everywhere denote the causal propagator without an index: $S(x,y)$.
Thus for $y^0=x^0+0$ we have (see Eq. (\ref{Torsion_Defin}))
\begin{gather}
\big\langle\overline{\Psi}_{\beta}(x+0)\Psi_{\alpha}(x)\big\rangle=-iS(x,x+0)_{\alpha\beta},
\nonumber \\
\mT_{abc}=-\frac{il^2_P}{4}\varepsilon_{abcd}\tr\gamma^5\gamma^dS(x,x+0).
\label{Average_through_Propagator}
\end{gather}
Since geometric quantities are assumed to be classical, we always assume $\mT_{abc}=\langle\mT_{abc}\rangle$.

Further, to simplify calculations, we use normal coordinates $x^{\mu}$ centered at point $p$, so that
$x^{\mu}(p)=0$ and in the vicinity of this point
\begin{gather}
e^a_{\mu}(x)=\delta^a_{\mu}+\frac12{\mT}^a_{\nu\mu}(p)x^{\nu}
\nonumber \\
+\left(\frac16{\mR}^a_{\nu\lambda\mu}+\frac16{\mT}^a_{\nu c}{\mT}^c_{\lambda\mu}+
\frac13{\mT}^a_{\nu\mu;\lambda}\right)(p)x^{\nu}x^{\lambda}
\nonumber \\
+\left(\frac{1}{24}\left[{\mR}^c_{\nu\lambda\mu}{\mT}^a_{\rho c}+
{\mR}^a_{\rho\nu c}{\mT}^c_{\lambda\mu}
+{\mT}^a_{\rho c}{\mT}^c_{\nu d}{\mT}^d_{\lambda\mu}\right]+\frac{1}{12}{\mR}^a_{\rho\nu\mu;\lambda} \right.
\nonumber \\
\left.
+\frac{1}{12}{\mT}^a_{\rho c}{\mT}^c_{\nu\mu;\lambda}
+\frac18{\mT}^a_{\rho c;\lambda}{\mT}^c_{\nu\mu} +\frac18{\mT}^a_{\rho\mu;\nu;\lambda}\right)(p)x^{\nu}x^{\lambda}x^{\rho},
\label{Tetrad_Normal_Coordinates}
\end{gather}
\begin{gather}
\omega_{ab\mu}(x)=\frac12\mR_{ab\nu\mu}(p)x^{\nu}
+\left(\frac16\mR_{ab\nu f}\mT^f_{\lambda\mu}
+\frac13\mR_{ab\nu\mu;\lambda}\right)(p)x^{\nu}x^{\lambda}.
\label{Connection_Normal_Coordinates}
\end{gather}
Using the formula (\ref{Tetrad_Normal_Coordinates}) we get:
\begin{gather}
\tilde{e}^{\mu}_a(x)=\delta^{\mu}_a+\delta \tilde{e}^{\mu}_a(x),
\nonumber \\
\delta \tilde{e}^{\mu}_a(x)=-\frac12\mT^{\mu}_{\nu a}(p)x^{\nu}+{\mE}^{\mu}_{\nu\lambda a}x^{\nu}x^{\lambda}
+{\mE}^{\mu}_{\nu\lambda\rho a}x^{\nu}x^{\lambda}x^{\rho},
\nonumber \\
{\mE}^{\mu}_{\nu\lambda a}=\left(-\frac16\mR^{\mu}_{\nu\lambda a}
+\frac{1}{12}\mT^{\mu}_{\nu c}\mT^c_{\lambda a}-\frac13\mT^{\mu}_{\nu a;\lambda}\right)(p),
\nonumber \\
{\mE}^{\mu}_{\nu\lambda\rho a}=\left(\frac{1}{24}\mR^c_{\nu\lambda a}{\mT}^{\mu}_{\rho c}+
\frac{1}{24}\mR^{\mu}_{\nu\lambda c}{\mT}^c_{\rho a}-\frac{1}{12}\mR^{\mu}_{\nu\lambda a;\rho}\right.
\nonumber \\
\left.+\frac{1}{12}{\mT}^{\mu}_{\nu c}{\mT}^c_{\lambda a;\rho}+\frac{1}{24}{\mT}^{\mu}_{\nu c;\rho}{\mT}^c_{\lambda a}-\frac18{\mT}^{\mu}_{\nu a;\lambda;\rho}\right)(p).
\label{ONB_Normal_Coordinates}
\end{gather}
In quantities (\ref{ONB_Normal_Coordinates})-(\ref{Connection_Normal_Coordinates}), the terms  of a higher degree relative to the coordinates $x^{\mu}$ will not be needed further.
Note that at the point $p$ Latin indices $a,b,\ldots$ and Greek indices $\mu,\nu,\ldots$ are indistinguishable.
Further, the argument $(p)$ for geometric quantities is omitted, since this does not lead to misunderstandings.

We will calculate the Dirac causal propagator in (\ref{Average_through_Propagator}) using perturbation theory.
Thus\footnote{Strictly speaking, $S(0,y)=\left\{\sqrt{-g}\left(i\tilde{e}^{\mu}_a\gamma^a{\cal D}_{\mu}-m\right)\right\}^{-1}_{0,y}=\left\{i\tilde{e}^{\mu}_a\gamma^a{\cal D}_{\mu}-m\right\}^{-1}_{0,y}(-g)^{-1/2}_{y}$, since $(-g)^{-1/2}_{z,y}\propto\delta^4(z-y)$.
But when $y\longrightarrow+0$ $g(y)\longrightarrow 1$, and we return to Eq. (\ref{Propagator_Expansion}).}
\begin{gather}
S(0,y)=\left\{ i\tilde{e}^{\mu}_a\gamma^a{\cal D}_{\mu}-m\right\}^{-1}_{0,y}=\{\left\{(i\gamma^{\mu}\partial_{\mu}-m)+V \right\}^{-1}_{0,y}
\nonumber \\
=S^{(0)}(-y)
-\int\d^4xS^{(0)}(-x)V(x)S^{(0)}(x-y)+\ldots,
\nonumber \\
S^{(0)}(x)=\int\frac{\d^4k}{(2\pi)^4}e^{-ikx}\frac{\gamma^{\mu}k_{\mu}+m}{k^2-m^2+i0},
\label{Propagator_Expansion}
\end{gather}
where (in the absence of an electromagnetic field)
\begin{gather}
V=i\delta\tilde{e}^{\mu}_a\gamma^a\partial_{\mu}+i\gamma^{\mu}\omega_{\mu}
+i\delta\tilde{e}^{\mu}_a\gamma^a\omega_{\mu}
\nonumber \\
=V^{(0)}+V^{(1)}+V^{(2)}.
\label{Perturbation}
\end{gather}
On the right-hand side of the last equality, $V^{(s)},  s=0,1,2$, denotes the contribution to the perturbation operator of degree $s$ relative to the coordinates $x^{\mu}$. The contributions of the powers $s>2$ are not interesting here, since they do not lead to divergent corrections. This implies that the expansions
(\ref{ONB_Normal_Coordinates})-(\ref{Connection_Normal_Coordinates}) relative to $x^{\mu}$ are correct, since the divergent integrals saturate at $x^{\mu}\longrightarrow0$.

Let's write out all $V^{(s)}$, taking into account that the degree of the operator $\partial_{\mu}$ is equal to $(-1)$. Using (\ref{ONB_Normal_Coordinates})-(\ref{Connection_Normal_Coordinates}) and (\ref{Perturbation}) we find:
\begin{gather}
V^{(0)}(x)=-\frac{i}{2}\gamma^a{\mT}^{\mu}_{\nu a}x^{\nu}\partial_{\mu},
\label{Null}
\end{gather}
\begin{gather}
V^{(1)}=i{\mE}^{\mu}_{\nu\lambda a}x^{\nu}x^{\lambda}\gamma^a\partial_{\mu}+
\frac{i}{4}{\mR}_{ab\mu\nu}x^{\mu}\gamma^{\nu}\sigma^{ab},
\label{First}
\end{gather}
\begin{gather}
V^{(2)}=i{\mE}^{\mu}_{\nu\lambda\rho a}x^{\nu}x^{\lambda}x^{\rho}\gamma^a\partial_{\mu}
+\frac{i}{2}\left(-\frac{1}{12}{\mR}_{bc\nu f}{\mT}^f_{\lambda a}+\frac13{\mR}_{bc\nu a;\lambda}\right)
x^{\nu}x^{\lambda}\gamma^a \sigma^{bc}.
\label{Second}
\end{gather}

\section{Calculations}

Obviously, the term of degree zero relative to $V$ on the right-hand side of Eq. (\ref{Average_through_Propagator}) is equal to zero. Indeed, we have
$\tr\gamma^5\gamma^dS^{(0)}(-y)\equiv0$. \\

\subsection{The contribution of the electromagnetic field to the torsion tensor}

Let us first calculate the contribution to the torsion from the electromagnetic field in the first order and without taking into account gravity. In this case  $V=e\gamma^{\mu}A_{\mu}$. According to (\ref{Average_through_Propagator}) and
(\ref{Propagator_Expansion}), this contribution is
\begin{gather}
\delta^{(1)}_A{\mT}_{abc}=\frac{i}{4}el^2_P\varepsilon_{abcd}\tr\gamma^5\gamma^d
\gamma^{\nu}\gamma^{\mu}\gamma^{\lambda}
\nonumber \\
\times\int\d^4x\left(\partial_{\nu}D^{(0)}(-x)\right)\left(\partial_{\lambda}D^{(0)}(x)\right)A_{\mu}(x)\equiv0,
\label{A_Contribution}
\end{gather}
since $D^{(0)}(-x)=D^{(0)}(x)$. Here $D^{(0)}(x)$ is the causal propagator of a free Boson field.

Comparison of Eqs. (\ref{Torsion_sim_Spin.Current}) and (\ref{A_Contribution}) shows that the mean  of the spin current in the canonical representation (\ref{Spin_Current}) in an external electromagnetic field is equal to zero in the first order in the field.
Note an interesting fact: a similar mean of the spin current, constructed using the representation of the spin operator in the Pauli-Lubanski form, is not equal to zero in an external electromagnetic field \cite{chu2021induced}.
Consequently, these two representations of spin current differ fundamentally in quantum field theory.

\subsection{Self-consistent equation for the torsion tensor}

Everywhere below, we assume that the electromagnetic field is zero.

{\bf 1.} Let us calculate the contribution to the right-hand side of Eq. (\ref{Average_through_Propagator}) in the first order in $V$.

a) Contribution from $V^{(0)}$:
\begin{gather}
\delta_{V^{(0)}}{\mT}_{abc}=-\frac34l^2_P\int_E\frac{\d^4k}{(2\pi)^4}\frac{k^2_E}{(k^2_E+m^2)^2}\cdot\mT_{abc}(p),
\nonumber \\
k^2_E=(k^1)^2+(k^2)^2+(k^3)^2+(k^4)^2.
\label{V_1_0_Contribution}
\end{gather}
Hereinafter, the Wick rotation is used to bring the integrals to a convenient form.

The squarely diverging integral in (\ref{V_1_0_Contribution}) must be cut off at a scale of the order of the Planck. This means that the maximum possible momentum is of the order
\begin{gather}
k_{\mbox{max}}\sim\frac{2\pi}{l_P}.
\label{Regularization}
\end{gather}
Therefore we have
\begin{gather}
\left(C_{V^{(0)}}\right)^2\equiv\frac34l^2_P\int_E^{k_{\mbox{max}}}\frac{\d^4k}{(2\pi)^4}\frac{k^2_E}{(k^2_E+m^2)^2}\sim 1.
\label{Estimation}
\end{gather}
\\

b) Contribution from $V^{(1)}$.

It is easy to see that this contribution is zero.
Indeed, such a contribution is proportional to integrals of the form
\begin{gather}
\int_E\frac{\d^4k}{(2\pi)^4}\frac{k^{\mu}}{(k^2_E+m^2)^m}=0,
\nonumber \\
\int_E\frac{\d^4k}{(2\pi)^4}\frac{k^{\mu}k^{\nu}k^{\lambda}}{(k^2_E+m^2)^{m+1}}=0,
\label{Zeroing}
\end{gather}
which are equal to zero after integration over the angles. Thus
\begin{gather}
\delta_{V^{(1)}}^{(1)}{\mT}_{abc}=0.
\label{V_1_1_Contribution}
\end{gather}

Further, we take into account only logarithmically divergent contributions, which are obtained by taking into account the potential $V^{(2)}$ in the first order, as well as $V^{(0)}\otimes V^{(1)}$ in the second and $V^{(0)}\otimes V^{(0)}\otimes V^{(0)}$ in the third order in $V$. In this case, taking into account (\ref{Regularization}), we have
\begin{gather}
\int_E^{k_{\mbox{max}}}\frac{\d^4k}{(2\pi)^4}\frac{1}{(k^2_E+m^2)^2}=\frac{1}{8\pi^2}\ln\frac{2\pi}{l_Pm},
\nonumber \\
\ln\frac{2\pi}{l_Pm}=\ln\frac{2\pi\hbar}{l_Pmc}\sim 50 \quad \mbox{for electron mass}.
\label{Estimate}
\end{gather}

Note that the estimate (\ref{Estimate}) is valid in the present epoch. We adhere to the version that the theory is regularized with the help of a lattice. This means that in the era close to the moment of the Big Bang, the volume of momentum space was larger in the same proportion as the volume of space decreased.
For this reason, near the Big Bang, we also have $k_{\mbox{max}}\longrightarrow\infty$, and the estimate (\ref{Estimate}) turns out to be much larger.

The self-consistent equation for torsion is written in terms of 4-vector ${\mt}^a$, which in the case under consideration is equivalent to the torsion tensor according to the equality
\begin{gather}
{\mT}_{abc}=\varepsilon_{abcd}{\mt}^d.
\label{Equivalence}
\end{gather}

In the process of calculations, the Bianchi identities are used, which have the following form in the presence of torsion:
\begin{gather}
{\mR}_{a[bcd]}={\mT}_{a[bc;d]}+{\mT}_{af[b}{\mT}^f_{cd]},
\label{Bianchi_1}
\end{gather}
\begin{gather}
{\mR}_{ab[cd;f]}=-{\mR}_{abe[f}{\mT}^e_{cd]}.
\label{Bianchi_2}
\end{gather}
Everywhere we use the standard notation for any multi-index value $\varkappa$: $\varkappa^{\ldots}_{\ldots[abc]}\equiv(\varkappa^{\ldots}_{\ldots abc}+\varkappa^{\ldots}_{\ldots bca}+\varkappa^{\ldots}_{\ldots cab})$. It is easy to check that, due to the representation (\ref{Equivalence}), the second term on the right-hand side of the equality (\ref{Bianchi_1}) vanishes identically:
${\mT}_{af[b}{\mT}^f_{cd]}\equiv0$. With (\ref{Bianchi_1}) we get:
\begin{gather}
{\mR}_{abcd}-{\mR}_{cdab}=\frac12\left({\mT}_{bcd;a}-{\mT}_{cda;b}-{\mT}_{dab;c}+{\mT}_{abc;d}\right).
\label{Broken_symmetry}
\end{gather}

The Ricci tensor by definition
\begin{gather}
{\mR}_{ab}\equiv\eta^{cd}{\mR}_{cadb}=\eta^{cd}{\mR}_{acbd}, \quad {\mR}_{ab}-{\mR}_{ba}={\mT}^c_{ba;c}.
\label{Ricci}
\end{gather}

c) As a result of cumbersome calculations, we get:
\begin{gather}
\delta_{V^{(2)}}{\mt}^a=
\frac{l_P^2}{96\pi^2}\ln\frac{2\pi}{l_Pm}\left\{3({\mt}^a)^{;b}_{;b}-3\left(\mt^b_{;b}\right)^{;a}
+\frac12{\mR}\cdot \mt^a-2{\mR}^a_{\,\,b}\mt^b+\varepsilon^{abcd}\mt_b\mt_{c;d}\right\}.
\label{V^{(2)}}
\end{gather}

{\bf 2.}  Contribution from $V\otimes V$:
\begin{gather}
\delta_{V\otimes V}{\mt}^a=
\frac{l_P^2}{96\pi^2}\ln\frac{2\pi}{l_Pm}\left\{
-\frac12{\mR}\cdot\mt^a-{\mR}^a_{\,\,b}\mt^b-\frac32\mt^2\cdot\mt^a+6\varepsilon^{abcd}\mt_b\mt_{c;d}\right\}.
\label{VV}
\end{gather}

{\bf 3.} Contribution from $V\otimes V\otimes V$:
\begin{gather}
\delta_{V\otimes V\otimes V}{\mt}^a=\frac{l_P^2}{96\pi^2}\ln\frac{2\pi}{l_Pm}\left\{
\frac32\mt^2\cdot\mt^a\right\}.
\label{VVV}
\end{gather}

{\bf 4.} Putting together the contributions
(\ref{Average_through_Propagator}), (\ref{V_1_0_Contribution}), (\ref{V_1_1_Contribution}),
(\ref{V^{(2)}}), (\ref{VV}) and (\ref{VVV}), we arrive at a self-consistent equation for the torsion tensor:
\begin{gather}
-\left[1+\left(C_{V^{(0)}}\right)^2\right]{\mt}^a+\frac{l_P^2}{96\pi^2}\ln\frac{2\pi}{l_Pm}
\left\{3(\mt^a)^{;b}_{;b}-3\left(\mt^b_{;b}\right)^{;a}
-3{\mR}^a_{\,\,b}\mt^b
+7\varepsilon^{abcd}\mt_b\mt_{c;d}\right\}=0.
\label{Self-consistent_Eq}
\end{gather}

It is useful to rewrite Eq. (\ref{Eq}) in the following form:  
\begin{gather}
e^{\mu}_be^{\nu}_c\left({\cal D}_{\mu}e_{a\nu}-{\cal D}_{\nu}e_{a\mu}\right)=\mT_{abc}
\equiv\varepsilon_{abcd}\mt^d.
\label{Eq_Mod}
\end{gather}

Some considerations about the possibility of the existence of non-zero solutions of Eq. (\ref{Self-consistent_Eq}) are given in the next Section.

\section{Discussion}

It would be interesting vacuum average the bilinear form of the Dirac fields in the Einstein equation $\delta\mA/\delta e^a_{\mu}=0$ in a similar way. Thus, the Einstein equation would arise, in which the contribution of the Dirac field to the matter energy-momentum tensor  would be expressed in terms of the curvature and torsion tensors. In deriving this equation, there would be additional physical difficulties: it would be necessary to make a subtraction from the cosmological constant, which diverges as the fourth power of the cutoff momentum and is due to the Dirac sea. This renormalization should lead to a very small cosmological constant. There was no such difficulty in deriving of Eq. (\ref{Self-consistent_Eq}). Another essential difference between the Einstein equation thus obtained and Eq. (\ref{Self-consistent_Eq}) is as follows. The energy-momentum tensor of real particles and antiparticles does not vanish if the curvature and torsion tensors are equal to zero (outside the mass shell).
Therefore, in the presence of real particles and antiparticles, the curvature tensor cannot vanish due to the Einstein equation (on the mass shell). The statement remains valid for zero torsion tensor.
On the contrary, Eq. (\ref{Self-consistent_Eq}) allows zeroing of the torsion tensor.

Let's assume that we have explicitly described Einstein equation.
In this case, we have a closed system of equations for variables $\mt^a$, $e^a_{\mu}$ and $\omega^{ab}_{\mu}$: this is Einstein equation  and Eqs. (\ref{Self-consistent_Eq}), (\ref{Eq_Mod})). To do this, it is necessary to express all the geometric quantities in Einstein equation in terms of the indicated variables. However, in this paper, Einstein equation (in the specified context) is not studied.

A step in this direction was taken in \cite{nascimento2022induced}: the topological correction to the Hilbert-Einstein-Cartan action was calculated, which is contained in the integral over the Dirac field in the one-loop approximation. However, this problem remains unresolved in its entirety at present.

According to the equation (\ref{Self-consistent_Eq}), the torsion field cannot propagate in space-time like the field of scalar (or vector) particles. This statement is not a new result, it can be found in the excellent review of Hehl et al \cite{hehl1976general}.
To clarify this statement, consider the Klein-Gordon-Fock equation for a scalar field in Minkowski space-time
\begin{gather}
(\partial_b\partial^b\phi+\mu^2\phi)=0
\label{Klein-Gordon-Fock}
\end{gather}
and an elementary consequence of this equation. In the case of a plane monochromatic wave,
$\phi\propto\exp(-ikx)$, Eq. (\ref{Klein-Gordon-Fock}) has a nonzero solution only if
\begin{gather}
k^0=\pm\sqrt{\mu^2+{\bf k}^2}.
\label{Mass_shell}
\end{gather}
In the case of a scalar particle, we have $\mu^2>0$. Therefore, scalar particles can propagate in space-time.

However, in Eq. (\ref{Self-consistent_Eq}) the role of the "squared mass" is played by the constant
\begin{gather}
\mu^2=-32\pi^2l_P^{-2}\left(\ln\frac{2\pi}{l_Pm}\right)^{-1}\left[1+\left(C_{V^{(0)}}\right)^2\right]<0.
\label{Mass_squared}
\end{gather}
Indeed, consider the case $\mt\longrightarrow0$ and $\mR^a_{\ \ bcd}\longrightarrow0$.
In this case, one can choose a gauge such as $\omega_{\mu}\longrightarrow0$ (see (\ref{Connection_Normal_Coordinates}), so that ${\cal D}_{\mu}\longrightarrow\partial_{\mu}$.
As a result, we can omit the last two terms in Eq. (\ref{Self-consistent_Eq}), 
and this equation breaks down into the following two equations:
\begin{gather}
\left(\partial_b\partial^b\mt^a
-32\pi^2l_P^{-2}\left(\ln\frac{2\pi}{l_Pm}\right)^{-1}\left[1+\left(C_{V^{(0)}}\right)^2\right]\right)=0,
\label{Self-consistent_Eq_Lin}
\end{gather}
and $\partial_b\mt^b=0$.  Comparison of Eqs. (\ref{Klein-Gordon-Fock}) and (\ref{Self-consistent_Eq_Lin}) leads to equality (\ref{Mass_squared}).
But then for ${\bf k}^2\ll |\mu^2|$ we have $k^0=-i|\mu|$, and the wave decays exponentially in time:
$\mt^a\propto\exp(-|\mu|x^0)$. Thus, only homogeneous stationary solutions or those close to them can be nonzero.

Such a situation (with an inverted mass square) takes place in Landau's theory of second-order phase transitions
below critical temperature.

There is also a significant difference between Eq. (\ref{Self-consistent_Eq}) and the corresponding equation in Landau's theory: in Eq. (\ref{Self-consistent_Eq}) there is no cubic term, but there is a quadratic term containing the first derivative. Obviously, Eq. (\ref{Self-consistent_Eq}) always has a zero solution for the torsion tensor. Finding nonzero solutions to this equation will be the subject of further research.
Here we only point out the possible existence of a certain solution. Let $\mt^a\longrightarrow0$ and gradient terms are unimportant,
and we will expand the solution in this small quantity. Then in the equations (\ref{Bianchi_1}) and (\ref{Bianchi_2}) the right parts can be set equal to zero in the leading approximation, and the curvature tensor can be taken in the form:
${\mR}^{ab}_{cd}=-H^2\left(\delta^a_c\delta^b_d-\delta^a_d\delta^b_c\right)$ (curvature of de Sitter space).
Then ${\mR}^a_b=-3H^2\delta^a_b$, and in the case $\mt^a_{;b}=0$ Eq. (\ref{Self-consistent_Eq}) is rewritten as
\begin{gather}
\left(\frac13\mu^2+H^2\right)\mt^a=0.
\label{Simplification}
\end{gather}
Here $\mu^2$ is given in (\ref{Mass_squared}). An equation (\ref{Simplification}) can have a non-zero solution if the parenthesis in that equation vanishes.
This effect is achieved with a certain and sufficiently large value of the parameter
\begin{gather}
H^2=\frac{32\pi^2}{3l_P^2}\left(\ln\frac{2\pi}{l_Pm}\right)^{-1}\left[1+\left(C_{V^{(0)}}\right)^2\right]\ll\frac{4\pi^2}{l_P^2}\sim k_{\mbox{max}}^2.
\nonumber
\end{gather}
This superficial consideration shows that if there are no conditions for the existence of the torsion tensor at the present epoch, then such conditions could take place near the Big Bang point.

Finally, we point out that, according to (\ref{Torsion_Defin}), we have the relation
\begin{gather}
J^{5a}=\left(4/l_P^2\right)\mt^a, \quad J^{5a}\equiv\overline{\Psi}\gamma^5\gamma^a\Psi.
\label{Acsial_Current}
\end{gather}
Equation (\ref{Acsial_Current}) shows that if the torsion tensor is not equal to zero, then the axial current is also not equal to zero. The physical consequences following from this are not clear to the author.

\begin{acknowledgments}

I thank Prof. G.E. Volovik for drawing my attention to the problem, as well as for useful advice in the process of work. This work was carried out as a part of the State Program 0033-2019-0005.

\end{acknowledgments}


\end{document}